\documentclass[letterpaper]{article} % DO NOT CHANGE THIS
\usepackage{aaai25}  % DO NOT CHANGE THIS
\usepackage{times}  % DO NOT CHANGE THIS
\usepackage{helvet}  % DO NOT CHANGE THIS
\usepackage{courier}  % DO NOT CHANGE THIS
\usepackage[hyphens]{url}  % DO NOT CHANGE THIS
\usepackage{graphicx} % DO NOT CHANGE THIS

\usepackage{algorithm}
\usepackage{algorithmic}
\usepackage{booktabs}
\usepackage{xcolor}
\usepackage{amsmath}

\usepackage{natbib}  % DO NOT CHANGE THIS AND DO NOT ADD ANY OPTIONS TO IT
\usepackage{caption} % DO NOT CHANGE THIS AND DO NOT ADD ANY OPTIONS TO IT
\usepackage{algorithm}
\usepackage{algorithmic}

\usepackage{newfloat}
\usepackage{listings}
\DeclareCaptionStyle{ruled}{labelfont=normalfont,labelsep=colon,strut=off} % DO NOT CHANGE THIS
\floatstyle{ruled}
\newfloat{listing}{tb}{lst}{}
\floatname{listing}{Listing}
\title{LLM-Fusion: A Novel Multimodal Fusion Model for \\ Accelerated Material Discovery}
\author {
    Onur Boyar\textsuperscript{\rm },
    Indra Priyadarsini\textsuperscript{\rm },
    Seiji Takeda\textsuperscript{\rm },
    Lisa Hamada\textsuperscript{\rm }
}
\affiliations {
    \textsuperscript{\rm }IBM Research - Tokyo\\
    Onur.Boyar@ibm.com, indra.ipd@ibm.com, seijitkd@jp.ibm.com,
    Lisa.Hamada@ibm.com
}

\usepackage{bibentry}

\begin{document}

\maketitle

\begin{abstract}
Discovering materials with desirable properties in an efficient way remains a significant problem in materials science. Many studies have tackled this problem by using different sets of information available about the materials. Among them, multimodal approaches have been found to be promising because of their ability to combine different sources of information. However, fusion algorithms to date remain simple, lacking a mechanism to provide a rich representation of multiple modalities. This paper presents LLM-Fusion, a novel multimodal fusion model that leverages large language models (LLMs) to integrate diverse representations, such as SMILES, SELFIES, text descriptions, and molecular fingerprints, for accurate property prediction. Our approach introduces a flexible LLM-based architecture that supports multimodal input processing and enables material property prediction with higher accuracy than traditional methods. We validate our model on two datasets across five prediction tasks and demonstrate its effectiveness compared to unimodal and naive concatenation baselines.
\end{abstract}

\section{Introduction}

In recent years, materials science has increasingly integrated AI-driven methods to accelerate the discovery of materials with specific properties. One prominent research direction focuses on the de novo generation of materials through generative models combined with optimization techniques \cite{gomez2018automatic, jin2018junction, boyar2024latent}. Another significant line of research aims to develop robust property prediction models, essential for screening large libraries of existing materials to identify those that exhibit desired characteristics \cite{soares2023beyond, soares2023improving, liu2023multi}. Both research directions involve unimodal and multimodal approaches. Unimodal approaches use a single representation, such as SMILES strings \cite{weininger_1988}, while multimodal architectures leverage various representations—such as SMILES, SELFIES \cite{krenn_2020}, graphs \cite{kajino2019molecular, kishimoto2023mhg}, fingerprints, and text descriptions. These diverse modalities are embedded and fused into a unified representation, enabling the model to effectively perform target tasks, whether reconstructing the original inputs or predicting material properties. In this paper, we focus on the latter—building a predictor model using a multimodal model via a novel fusion mechanism.
\vspace{-2.5ex}
\begin{figure}[ht]
    \centering
    \includegraphics[page=2,scale=0.37]{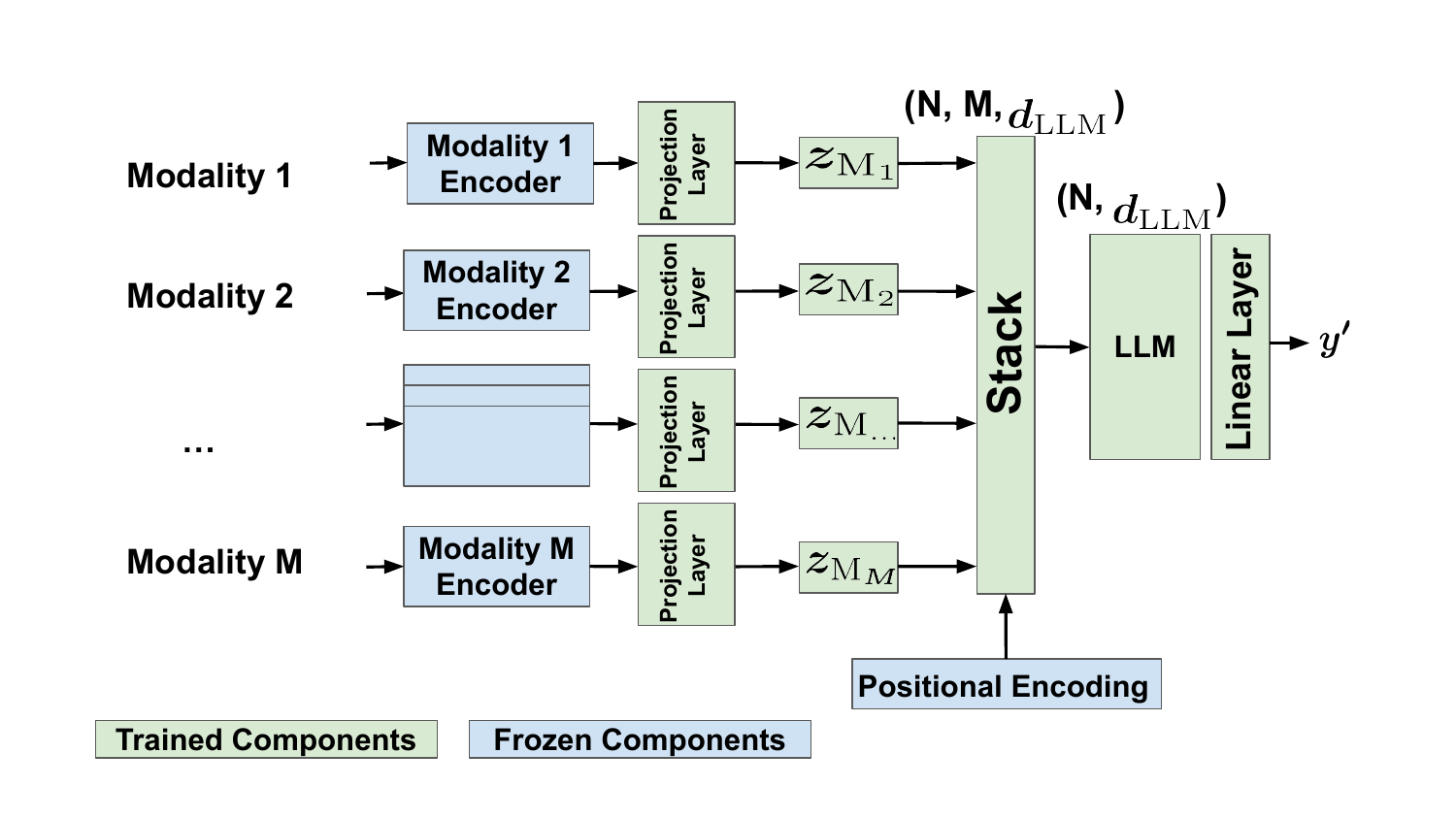}
    \vspace{-2ex}
    \caption{LLM-Fusion architecture.}
    \label{fig:architecture}
\end{figure}

A common approach to fusing different modalities is through \textit{naive concatenation} \cite{hu2016segmentation_vision_concat1, soares2023beyond}, which combines the embeddings of each modality into a single, concatenated vector. In property prediction tasks, these concatenated vectors often serve as inputs to predictor models. However, this approach has several limitations. One drawback is that it disregards the relationships among the different modalities. An alternative approach is \textit{contrastive learning}, which aims to structure the latent spaces of each modality such that embeddings from different modalities corresponding to the same material are located in similar regions. While this alignment can be effective, contrastive learning becomes increasingly challenging to scale beyond two modalities due to the exponential growth in the number of modality pairs and the computational complexity of managing numerous positive and negative samples during training \cite{takeda2023foundation}.

In this paper, we propose LLM-Fusion (see Fig. \ref{fig:architecture}), a novel fusion technique that leverages Large Language Models (LLMs) for multimodal modeling. LLMs are highly effective across diverse domains due to their ability to process complex, unstructured data, and also shown promise in numerical prediction tasks \cite{vacareanu2024words}. Our approach employs an LLM-based fusion model that efficiently scales with additional modalities, providing fixed-size, information-dense representations of multimodal inputs. This scalability makes LLM-Fusion particularly well-suited for property prediction tasks in materials science. We evaluate LLM-Fusion using combinations of two, three, and four modalities, offering, to our knowledge, the most extensive analysis of performance and scalability among multimodal property prediction models. Our experiments across two datasets and five prediction tasks demonstrate that predictive performance consistently improves as more modalities are included, showcasing the potential of LLM-Fusion.

%In this paper, we propose a novel fusion technique, LLM-Fusion (see Fig. \ref{fig:architecture}), which utilizes Large Language Models (LLMs) for multimodal modeling. LLMs have proven highly effective across diverse domains due to their ability to process complex, unstructured data, and they have recently shown promise in numerical prediction tasks as well \cite{vacareanu2024words}. Our approach leverages an LLM-based fusion model that scales efficiently with an increasing number of modalities, providing fixed-size, information-dense representations of multimodal inputs. This scalability makes LLM-Fusion particularly well-suited for property prediction tasks in materials science. We evaluate LLM-Fusion using combinations of two, three, and four modalities, providing, to our knowledge, the most extensive analysis of performance and scalability among multimodal property prediction models in the literature. Our experiments across two datasets and five prediction tasks demonstrate that predictive performance consistently improves with the inclusion of additional modalities, showcasing the potential of the LLM-Fusion approach.

%\vspace{-4ex}
\section{Related Work}
%\vspace{-1.2ex}
Traditional models in materials science often rely on unimodal representations, such as SMILES strings and molecular fingerprints, for property prediction. For example, ChemBERTa \cite{chithrananda2020chemberta} applies a BERT-like architecture to SMILES strings \cite{weininger_1988} for learning powerful representations of materials, which can be used for predicting material properties. Morgan fingerprints \cite{Morgan1965, Rogers2010}, which represent molecules as binary vectors indicating the presence or absence of specific substructures, are also widely used for property prediction. Another representation gaining popularity is SELFIES \cite{krenn_2020}. Studies like \citet{yuksel2023selformer, priyadarsini2024improving} propose architectures trained on a large corpus of SELFIES representations of materials.

However, multimodal approaches, which integrate multiple types of data, require more general mechanisms and are often inspired by methodologies developed outside the materials science domain. In computer vision, for example, the field of multimodal learning has been pioneered by models such as CLIP \cite{radford2021learning}, which employs contrastive learning in text-based image generation. Alternatively, many studies \cite{hu2016segmentation_vision_concat1, li2018referring_vision_concat2, shi2018key_vision_concat3} utilize concatenation of representations from different modalities to perform their target tasks. In the material science community, multimodal approaches are focused on both material generation tasks and property prediction tasks, such as in \citet{liu2023multi}, where they proposed a methodology for text-based editing of molecules, using a contrastive learning-based approach. Another recent approach is \citet{liu2024git}, which uses a cross-attention mechanism to learn the fused representations. In another example, \citet{soares2023beyond} utilizes graph-based representation and SMILES representation as two modalities and uses a naive concatenation-based approach for the property prediction task. Similarly, 
\citet{soares2023improving} uses the SMILES representation and tabular data containing a wide variety of molecular features as two modalities and uses a naive concatenation-based approach for the property prediction task. Among these models, we selected naive concatenation as our main competitor model because, unlike other alternatives, it easily scales beyond two modalities and has proven to be useful in many studies.

\section{Proposed Method}

In this section, we first discuss our motivation to use LLM as the fusion model, and then provide the details of the LLM-Fusion architecture.

\subsection{Motivation}
An effective multimodal fusion methodology should be flexible, seamlessly adapting as modalities are added or removed. However, existing studies often lack evaluations of their methods' adaptability to varying numbers of modalities and do not discuss their applicability in dynamic scenarios. Therefore, there is a need for a fusion model that can adjust to different numbers of modalities and improve its performance as new information becomes available.

LLMs are inherently good at summarization of given contents. They are equipped with powerful self-attention mechanisms that allow them to model complex dependencies. Besides, LLMs are the most popular models of recent literature and there is a rapid development of their capabilities. Motivated by these, we implemented them for a fusion task. By extending their capabilities to model the sequences of modality embeddings, LLM-Fusion leverages the LLM's ability to capture interactions between different types of modalities. The LLM's self-attention mechanism dynamically weighs the importance of each modality's features for the prediction task and fuses multiple modalities into a fixed-length, unified representation. Besides, since our proposed method does not focus on a specific type of LLM, as the capabilities of LLMs increase, their performance on the LLM-Fusion framework can also increase.

\subsection{LLM-Fusion}

Our LLM-Fusion model can work with $M$ modalities. The encoders for the each of the $M$ modalities can be of any architecture. LLM-Fusion uses the batch of $N$ 1-dimensional embeddings obtained by these $M$ unimodal encoders.  The encoders can be frozen if they are already pretrained, and can be trained from scratch if they are not. In the proposed LLM-Fusion model, we do not fine-tune the pretrained encoders. However, in the case of using pretrained encoders, the embedding size of the pretrained encoder and fusion models input embedding dimension may differ. For this reason, we introduced \textit{optional} projection layers. These are used to match the embedding dimension of each embedding with the input embedding dimension of LLM, $d_{\text{LLM}}$, if needed. In Fig. \ref{fig:architecture}, we assumed that all encoders are pretrained encoders, which may require projection layers to match the embedding dimension of the LLM-Fusion model.

Following the optional projection layer, the encoded vectors or transformed vectors via projection layers are stacked along a new dimension, forming a tensor of shape $(N, M, d_{\text{LLM}})$. The stacked tensor is enriched with positional encodings that specify the location of each modality within the tensor, providing information to distinguish between different sources of information by treating them as a sequence of modalities. This stacked and positional encoding added tensor serves as input to the LLM, which is fed at the input embeddings layer of the LLM. This strategy allows us to skip the traditional tokenization and positional encoding addition steps of transformer training. Instead, we provide positional encodings based on modalities and provide the embeddings directly from the embedding layer of the LLM. Using this input, the LLM-Fusion model outputs a fixed-size vector of fused representation used for property prediction, which has a dimension of $(N, d_{\text{LLM}})$. This vector is obtained by taking the average of the last hidden state of the LLM. Note that, even with the increased number of modalities, the size of the embedding vector remains the same. The final layer of our model is a single linear layer that uses the fused representation as input and provides the property prediction.

\begin{figure*}[h!]
    \centering
    % First figure
    \begin{minipage}{0.48\textwidth}
        \centering
        \includegraphics[page=4,scale=0.33]{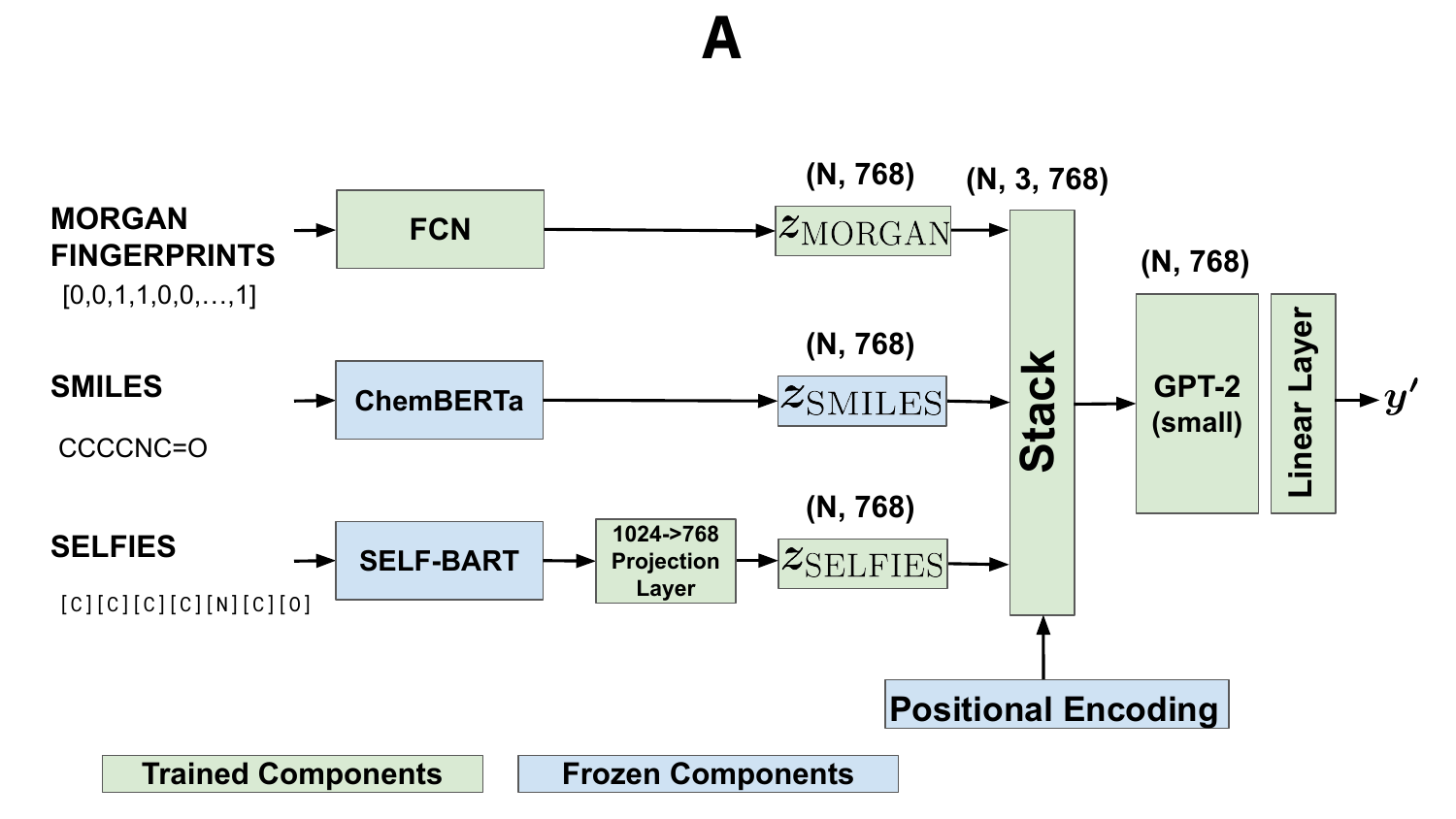}
        %\caption{Caption for the first figure}
        %\label{fig:first_figure}
    \end{minipage}
    \hfill
    % Second figure
    \begin{minipage}{0.48\textwidth}
        \centering
        \includegraphics[page=3,scale=0.33]{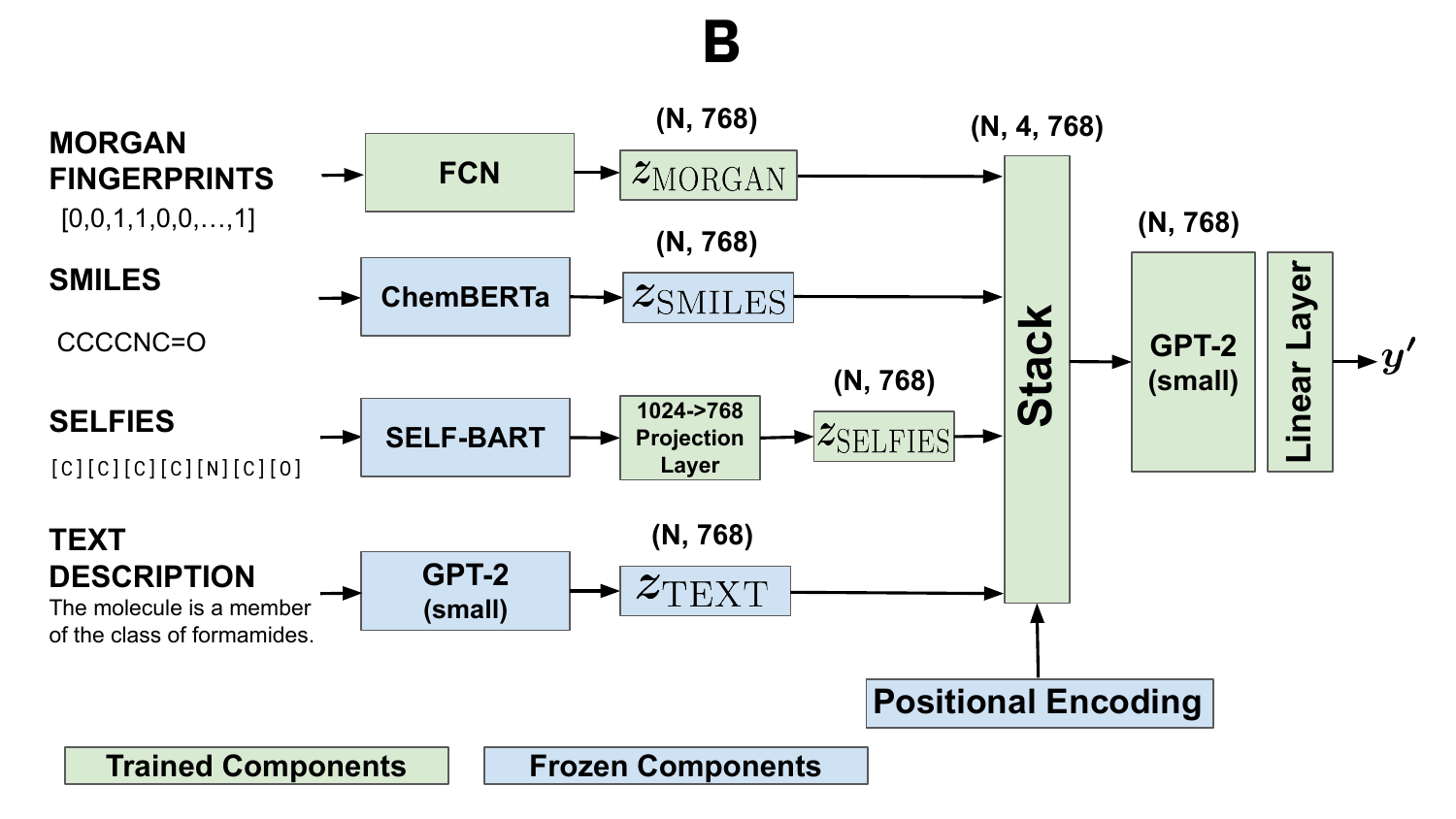}
        %\caption{Caption for the second figure}
        %\label{fig:second_figure}
    \end{minipage}
    \vspace{-1ex}
    \caption{LLM-Fusion models used in experiments. \textbf{A} shows the three-modality model used for the HOMO, LUMO, and GAP property prediction tasks, \textbf{B} shows the four-modality model used for the LogP and QED prediction tasks.}
    \label{fig:two_figures_side_by_side}
\end{figure*}
\vspace{-2ex}
\section{Experiments}

In this section, we discuss our experimental setup, training details, results, and performance improvement when a larger LLM is used as a fusion model.

\subsection{Datasets}

In our experiments, we used two different datasets. In the first set of experiments, we used the MoleculeNet QM9 dataset \cite{wu2018moleculenet}, which is a common benchmark dataset in property prediction tasks. Experiments using this dataset used SMILES, SELFIES, and Morgan fingerprints modalities, and subsets of them. We evaluated our model on HOMO, LUMO, and GAP property prediction tasks. 

The second dataset we used was the ChEBI-20 dataset proposed by \citet{edwards2022translation}, which includes SMILES strings of materials and their text descriptions. The availability of text descriptions enabled us to increase the number of modalities to four,  namely SMILES, SELFIES, Morgan fingerprints, and text descriptions. Using this dataset, we trained LLM-Fusion models with different combinations of these modalities and evaluated them in LogP and Quantitative Estimate of Drug-likeness (QED) property prediction tasks.
\vspace{-0.5ex}
\subsection{Fusion Model Selection and Modalities}

%In our experiments, we used different sets of modalities to see how the model scales as the number of modalities increases. Specifically, we trained models with two, three, and four modalities. 

As the fusion model, we chose GPT-2 Small \cite{radford2019language}, motivated by several factors. With only 114 million parameters, GPT-2 Small is considerably smaller than modern LLMs, making it computationally efficient and allowing for faster training and inference times—an important consideration when dealing with multiple modalities and large datasets. Despite its reduced size, it retains the powerful self-attention mechanisms inherent in LLMs, enabling it to capture complex dependencies and interactions between modalities while keeping computational demands manageable. This balance of efficiency and capability makes GPT-2 Small a practical choice for our fusion model.

In Fig. \ref{fig:two_figures_side_by_side}, we provide two example architectures that we used in our experiments. Fig. \ref{fig:two_figures_side_by_side}(A) shows the case where three modalities, Morgan fingerprints, SMILES, and SELFIES, are used, where (B) shows the case with four modalities where Morgan fingerprints, SMILES, SELFIES, and Text Descriptions are used. In both models, among the encoders, only the encoder for Morgan fingerprints is trained, while the remaining modality encoders' weights are kept frozen. In the case of ChemBERTa used for SMILES modality, its class token is used as the embedding to be provided to LLM-Fusion model, while for SELF-BART \cite{priyadarsini2024improving} used for SELFIES modality, and GPT-2 Small is used for the text description modality, the average of the last hidden state of the respective models are used as the embeddings. For the Morgan fingerprints encoder, we used a single linear layer followed by a ReLu activation function. As demonstrated in Fig. \ref{fig:two_figures_side_by_side}, the encoders for Morgan fingerprints, SMILES, and text descriptions encoder provides 768-dimensional embeddings, while SELF-BART provides 1024-dimensional embeddings. Since the embedding size of the LLM-Fusion model, GPT-2 Small, is also 768, we only needed to use a projection layer for the SELFIES modality.%which decreases the embedding dimension to 768 from 1024.

\subsection{Benchmark Models and Training Details}

We compare the performance of LLM-Fusion with unimodal models and naive concatenation-based approaches. Following previous studies \cite{soares2023beyond, soares2023improving}, we used XGBoost \cite{chen2016xgboost} as the predictor for the unimodal and concatenation cases. The naive concatenation method employs the same modalities as LLM-Fusion to evaluate performance as the number of modalities increases. XGBoost hyperparameters were selected via a random search based on the validation set performance\footnote{Tuned hyperparameters: \textit{learningRate, maxDepth, nEstimators, subSample, colSampleByTree, gamma, minChildWeight.}}. We used predefined training-validation-test splits for both the QM9 (108,446/12,050/13,389 instances) and ChEBI-20 (26,403/3,301/3,299 instances) datasets. LLM-Fusion was trained using the AdamW optimizer with weight decay and PyTorch's ReduceLROnPlateau learning rate scheduler based on validation loss. %Models are trained on a single GPU with 24 hours od computation time doe all of the models. 

\begin{table}[h!]
\centering
%\tiny
%\small
\begin{tabular}{lcccc}
\toprule
\textbf{Model} & \textbf{Modalities} & \textbf{HOMO} $\downarrow$ & \textbf{LUMO} $\downarrow$ & \textbf{GAP} $\downarrow$ \\
\midrule
 UM & MFP & 0.0076 & 0.0092 & 0.0113  \\
 UM & SM & 0.0102 & 0.0141 & 0.0154 \\  
 UM & SF  & 0.0160 & 0.030 & 0.030 \\    \hline
  NC & SM+MFP & 0.0074 & 0.0090 & 0.0113 \\
 NC & SF+MFP & 0.0076 & 0.0092 & 0.0113 \\
 NC & SM+SF+MFP & 0.0074 & 0.0090 & 0.0113 \\  \hline
 Ours & SM+MFP & 0.0063 & 0.0080 & 0.0085  \\
 Ours & SF+MFP & 0.0067 & 0.0080 & 0.0079  \\
 %LLM Fusion & SMILES + SELFIES & - & - & -  \\
 Ours & SM+SF+MFP & \textbf{\textcolor{red}{0.0053}} & \textbf{\textcolor{red}{0.0055}} & \textbf{\textcolor{red}{0.0078}}  \\
\bottomrule
\end{tabular}
\vspace{-1ex}
\caption{Results after fine-tuning the LLM-Fusion (Ours) model and benchmarking against naive concatenation (NC) + XGBoost and unimodal (UM) + XGBoost models. 
}
\label{tab:mse_summary_qed}
\end{table}

\begin{table}
\centering
%\tiny
\begin{tabular}{lccc}
\toprule
\textbf{Model} & \textbf{Modalities} & \textbf{LogP} $\downarrow$ & \textbf{QED} $\downarrow$ \\
\midrule
 UM & MFP  & 3.083 & 0.0091  \\
 UM & SM & 5.522 & 0.0111  \\ 
UM & TEXT & 9.131 & 0.0212  \\ 
UM & SF & 2.521 & 0.0077  \\ \hline
 NC & SM+TEXT & 4.971 & 0.0112 \\
 NC & SM+TEXT+MFP & 2.022 & 0.0061  \\  
 NC & SM+TEXT+MFP+SF & 2.002 & 0.0063  \\   \hline
 Ours & SM+TEXT & 2.781 & 0.0062  \\
Ours & SSM+TEXT+MFP & 1.578 & 0.0034 \\ 
Ours & SM+TEXT+MFP+SF & \textbf{\textcolor{red}{1.332}} & \textbf{\textcolor{red}{0.0029}} \\
\bottomrule
\end{tabular}
\vspace{-1ex}
\caption{Results after fine-tuning the LLM-Fusion (Ours) model and benchmarking against naive concatenation (NC) + XGBoost and unimodal (UM) + XGBoost models.
}
\label{tab:mse_summary_chebi}
\end{table}
\vspace{-4ex}
\subsection{Results}

Tables \ref{tab:mse_summary_qed} and \ref{tab:mse_summary_chebi} summarize the results. In each table, SMILES, Morgan fingerprints, SELFIES modalities denoted as SM, MFP, and SF, respectively.  Results are presented as Mean Absolute Error (MAE) in Table \ref{tab:mse_summary_qed} and as Mean Squared Error (MSE) in Table \ref{tab:mse_summary_chebi}. As can be seen from the table, we evaluated models with different combinations of modalities, for both our proposal and naive concatenation approach, and it is clear that as the number of modalities increases, the predictive performance of LLM-Fusion increases. On the other hand, we understand that the naive concatenation method fails to scale its performance with the increased number of modalities. One reason for this phenomenon for the naive concatenation approach is that the dimension of the input vector keeps increasing as new modalities are added, which causes the curse of dimensionality. On the other hand, our approach can enrich the fixed-size fused representation as new information is provided. 

A similar pattern is observed in Table \ref{tab:mse_summary_chebi} for the LogP and QED prediction tasks across the two, three, and four-modality settings. LLM-Fusion consistently outperforms both unimodal models and naive concatenation baselines, achieving lower MSE values for both tasks. At each combination of modalities, LLM-Fusion provides better performance than the naive concatenation method, and its performance improves with the addition of more modalities. This demonstrates that even though the fused representation dimension remains the same, adding new modalities enriches the representation and enhances predictive performance.

%\subsection{Ablation Study} 

%As an ablation study, we removed the positional encoding from the LLM-Fusion model and evaluated it on the 4-modality case on ChEBI-20 and the 3-modality case on the QM9 dataset.

%\begin{table}[h!]
%\centering
%\tiny
%\begin{tabular}{lcccc}
%\toprule
%\textbf{Modalities} & \textbf{Task} & w/o P.E. &  w. P.E.\\
%\midrule
%SM+SF+MFP & HOMO & - & - \\ 
%SM+SF+MFP &  LUMO & - & - \\ 
%SM+SF+MFP &  GAP & - & - \\ 
%SM+TEXT+MFP+SF & LOGP & - & - \\
%SM+TEXT+MFP+SF & QED & - & - \\
%\bottomrule
%\end{tabular}
%\caption{Ablation results.}
%\label{tab:ablation}
%\end{table}
\vspace{-1ex}
\subsection{Using a Larger LLM}

In our experiments, we used the GPT-2 Small model as the fusion model. We conducted a small study to see how the model behaves when it is replaced by a larger model. We replaced the GPT-2 Small model with the GPT-2 Large \cite{radford2019language} model, which has 784 million parameters, and evaluated it using the four-modality case in the ChEBI-20 dataset. GPT-2 Large has an embedding dimension of 1280, therefore, for each modality except Morgan fingerprints, we employed projection layers and trained the LLM-Fusion model for LogP and QED property prediction tasks. 

\begin{table}[h!]
\centering
%\tiny
\begin{tabular}{lcccc}
\toprule
\textbf{Modalities} & \textbf{LLM Type} & \textbf{LogP} $\downarrow$ & \textbf{QED} $\downarrow$ \\
\midrule
SM+TEXT+MFP+SF & GPT-2 Small & 1.332 & 0.0029 \\
SM+TEXT+MFP+SF & GPT-2 Large & \textbf{\textcolor{red}{1.270}} & \textbf{\textcolor{red}{0.0027}} \\
\bottomrule
\end{tabular}
\vspace{-1ex}
\caption{Prediction performance when GPT-2 Small model is replaced by GPT-2 Large model.}
\label{tab:gtp2_large}
\end{table}

Table \ref{tab:gtp2_large} shows our results. For a fair comparison, we trained the GPT-2 Large model using the same GPU time (24 hours) as the GPT-2 Small model, which resulted in a smaller number of training epochs for the GPT-2 Large model due to differences in model sizes. However, the GPT-2 Large model still provides better results compared to the GPT-2 Small model, supporting our claim of obtaining a better fusion model by simply replacing the LLM itself.
\vspace{-0.5ex}
\section{Conclusion and Future Work}
LLM-Fusion represents a novel approach to multimodal property prediction in material science by leveraging the capabilities of large language models. Our results demonstrate that this architecture can significantly enhance property prediction accuracy, streamline the integration of diverse data types, and provide a scalable solution for material discovery. Future work will explore the application of this model to molecular generation tasks, to be used in scenarios such as text-based editing of materials. Besides, we will also explore the impact of using the more recent family of LLMs such as LLaMa \cite{touvron2023llama} or Granite \cite{mishra2024granite} models as our fusion model. The major drawback of our proposal is the computational complexity of the LLM-Fusion model. For this reason, we also aim to work on a general LLM-Fusion model that is pretrained on a larger number of modalities using a larger corpus of materials, to be readily used in downstream tasks.

%\textbf{modeldeki linear layeri check et (OK), belki ismini regression head yapmak daha iyi olur, o gercekten linear mi her modelde? experimental setup kismi cok uzun, ne yapabilirim?}

\clearpage
\bibliography{aaai25}

\end{document}